\documentclass[twoside,twocolumn,9pt]{article}
\usepackage{extsizes}
\usepackage[super,sort&compress,comma]{natbib} 
\usepackage[version=3]{mhchem}
\usepackage[left=1.5cm, right=1.5cm, top=1.785cm, bottom=2.0cm]{geometry}
\usepackage{balance}
\usepackage{mathptmx}
\usepackage{sectsty}
\usepackage{graphicx} 
\usepackage{lastpage}
\usepackage[format=plain,justification=justified,singlelinecheck=false,font={stretch=1.125,small,sf},labelfont=bf,labelsep=space]{caption}
\usepackage{float}
\usepackage{fancyhdr}
\usepackage{fnpos}
\usepackage[english]{babel}
\addto{\captionsenglish}{%
  
}
\usepackage{array}
\usepackage{droidsans}
\usepackage{charter}
\usepackage[T1]{fontenc}
\usepackage[usenames,dvipsnames]{xcolor}
\usepackage{setspace}
\usepackage[compact]{titlesec}
\usepackage{hyperref}

\usepackage{epstopdf}

\definecolor{cream}{RGB}{222,217,201}

\begin{document}

\pagestyle{fancy}
\thispagestyle{plain}
\fancypagestyle{plain}{
\renewcommand{\headrulewidth}{0pt}
}

\makeFNbottom
\makeatletter
\renewcommand\LARGE{\@setfontsize\LARGE{15pt}{17}}
\renewcommand\Large{\@setfontsize\Large{12pt}{14}}
\renewcommand\large{\@setfontsize\large{10pt}{12}}
\renewcommand\footnotesize{\@setfontsize\footnotesize{7pt}{10}}
\makeatother

\renewcommand{\thefootnote}{\fnsymbol{footnote}}
\renewcommand\footnoterule{\vspace*{1pt}%
\color{cream}\hrule width 3.5in height 0.4pt \color{black}\vspace*{5pt}} 
\setcounter{secnumdepth}{5}

\makeatletter 
\renewcommand\@biblabel[1]{#1}            
\renewcommand\@makefntext[1]%
{\noindent\makebox[0pt][r]{\@thefnmark\,}#1}
\makeatother 
\renewcommand{\figurename}{\small{Fig.}~}
\sectionfont{\sffamily\Large}
\subsectionfont{\normalsize}
\subsubsectionfont{\bf}
\setstretch{1.125} 
\setlength{\skip\footins}{0.8cm}
\setlength{\footnotesep}{0.25cm}
\setlength{\jot}{10pt}
\titlespacing*{\section}{0pt}{4pt}{4pt}
\titlespacing*{\subsection}{0pt}{15pt}{1pt}

\fancyfoot{}
\fancyfoot[LO,RE]{\vspace{-7.1pt}\includegraphics[height=9pt]{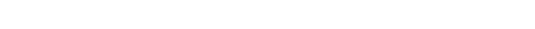}}
\fancyfoot[CO]{\vspace{-7.1pt}\hspace{13.2cm}\includegraphics{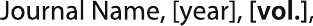}}
\fancyfoot[CE]{\vspace{-7.2pt}\hspace{-14.2cm}\includegraphics{head_foot/RF}}
\fancyfoot[RO]{\footnotesize{\sffamily{1--\pageref{LastPage} ~\textbar  \hspace{2pt}\thepage}}}
\fancyfoot[LE]{\footnotesize{\sffamily{\thepage~\textbar\hspace{3.45cm} 1--\pageref{LastPage}}}}
\fancyhead{}
\renewcommand{\headrulewidth}{0pt} 
\renewcommand{\footrulewidth}{0pt}
\setlength{\arrayrulewidth}{1pt}
\setlength{\columnsep}{6.5mm}
\setlength\bibsep{1pt}

\makeatletter 
\newlength{\figrulesep} 
\setlength{\figrulesep}{0.5\textfloatsep} 

\newcommand{\topfigrule}{\vspace*{-1pt}%
\noindent{\color{cream}\rule[-\figrulesep]{\columnwidth}{1.5pt}} }

\newcommand{\botfigrule}{\vspace*{-2pt}%
\noindent{\color{cream}\rule[\figrulesep]{\columnwidth}{1.5pt}} }

\newcommand{\dblfigrule}{\vspace*{-1pt}%
\noindent{\color{cream}\rule[-\figrulesep]{\textwidth}{1.5pt}} }

\makeatother

\twocolumn[
  \begin{@twocolumnfalse}
{\includegraphics[height=30pt]{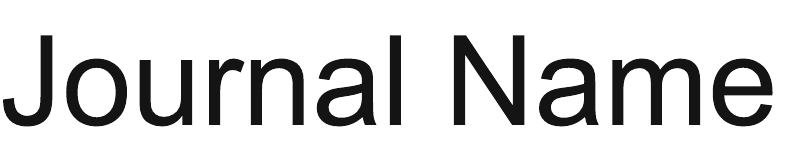}\hfill\raisebox{0pt}[0pt][0pt]{\includegraphics[height=55pt]{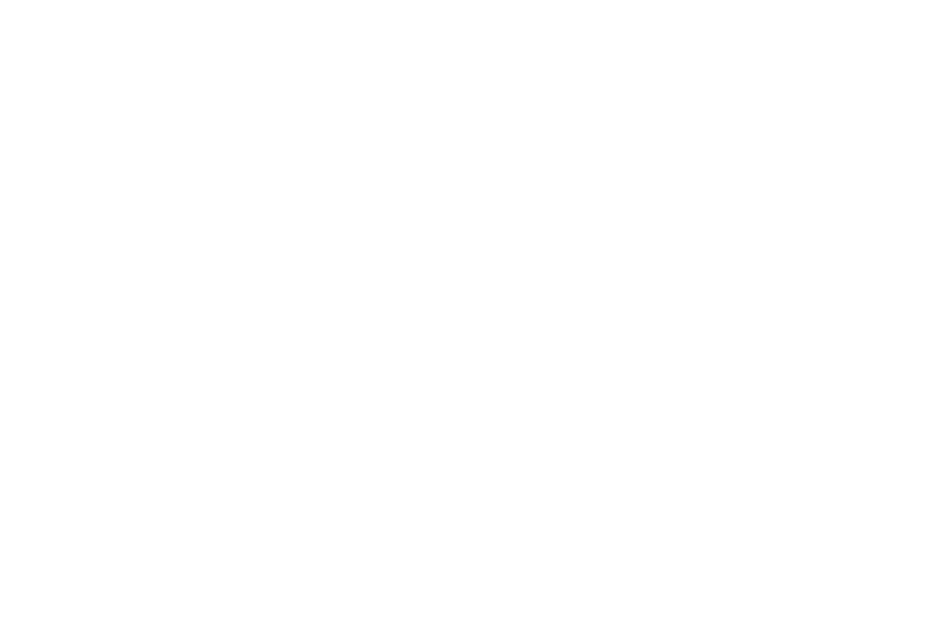}}\\[1ex]
\includegraphics[width=18.5cm]{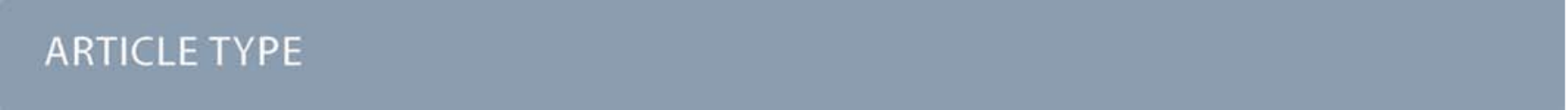}}\par
\vspace{1em}
\sffamily
\begin{tabular}{m{4.5cm} p{13.5cm} }

\includegraphics{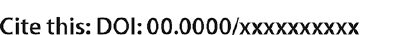} & \noindent\LARGE{\textbf{Interplay of structural and dynamical heterogeneity in the nucleation mechanism in Ni}} \\
\vspace{0.3cm} & \vspace{0.3cm} \\

 & \noindent\large{Grisell D\'{i}az Leines,$^{\ast}$\textit{$^{a}$} Angelos Michaelides,\textit{$^{a}$} and Jutta Rogal\textit{$^{b,c}$}} \\
 
\includegraphics{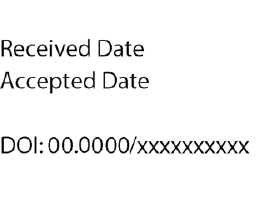} & \noindent\normalsize{ Gaining fundamental understanding of crystal nucleation processes in metal alloys is crucial for the development and design of high-performance materials with targeted properties. 
Yet,  crystallization is a complex non-equilibrium process and, despite having been studied for decades, the microscopic aspects that govern the crystallization mechanism 
of a material remain to date elusive. 
Recent evidence shows that spatial heterogeneity in the supercooled liquid, characterised by extended regions with distinctive mobility and order,
may be a key microscopic factor that determines the mechanism of crystal nucleation. These findings have advanced
our view of the fundamental nature of crystallization, as most research has assumed that crystal clusters nucleate from random fluctuations in a `homogeneous' liquid. Here, by analysing transition path sampling trajectories, we show that dynamical heterogeneity plays a key role in the mechanism of crystal nucleation in an elemental metal, nickel. 
Our results demonstrate that crystallization occurs preferentially in regions of low mobility in the supercooled liquid, evidencing the collective dynamical nature of crystal nucleation in Ni. 
In addition, our results show that low mobility regions form before and spatially overlap with pre-ordered domains that act as precursors to the crystal phase that subsequently emerges.
Our results show  a clear link between dynamical and structural heterogeneity in the supercooled liquid and its impact on the nucleation mechanism, revealing microscopic descriptors that could pave a novel way to control crystallization processes in metals. }

\end{tabular}

 \end{@twocolumnfalse} \vspace{0.6cm}

  ]

\renewcommand*\rmdefault{bch}\normalfont\upshape
\rmfamily
\section*{}
\vspace{-1cm}


\footnotetext{\textit{$^{a}$~Yusuf Hamied Department of Chemistry, University of Cambridge, Lensfield Road, Cambridge CB2 1EW, United Kingdom. E-mail: gd466@cam.ac.uk}}
\footnotetext{\textit{$^{b}$~Department of Chemistry, New York University, New York, NY 10003, USA. E-mail: jutta.rogal@nyu.edu }}
\footnotetext{\textit{$^{c}$~Fachbereich Physik, Freie Universit{\"a}t Berlin, 14195 Berlin, Germany}}



\section{Introduction}

Crystal nucleation is one of the most fundamental physical processes, relevant to a large range of everyday phenomena, from cloud formation to drug development. Yet, despite being studied for decades, 
it is still not well understood at the microscopic level.~\cite{Sosso2016,doi:10.1063/5.0055248} During homogeneous nucleation, the formation of a crystal cluster 
is induced by
fluctuations of competing translational and orientational orders in the supercooled liquid.~\cite{PhysRevX.8.021040} 
It has been assumed so far that a nucleus emerges from random fluctuations of order in a `homogeneous' liquid state. However, several experimental and computational studies have revealed that liquids can exhibit much more complex behaviour, such as structural and dynamical heterogeneity,~\cite{Gebauer2014,tenWolde1997,tenWolde1999,Zhang2007,PhysRevLett.108.225701,Russo2012,Fitzner2019,PhysRevX.8.021040}  where spatial regions with different structural features and mobility coexist in the liquid. Recently, it has been shown that the heterogeneities in the liquid are highly correlated with the nucleation ability of a material and the polymorph formed, revealing the fascinating collective behaviour of supercooled liquids and its connection to crystallization mechanisms.~\cite{DLeines2018,Tanaka2012, PhysRevX.8.021040,Russo2012,DLeines2017,Fitzner2019,Zhang2019, Lechner2011a,Tan2014,PhysRevLett.105.025701,PhysRevLett.96.175701} 
Yet, our fundamental understanding of the  nature of supercooled liquids is far from complete. Further studies are required to establish the general connections between liquid characteristics and crystal nucleation mechanisms in order to derive novel fundamental rules and microscopic descriptors. 
Not only are such insights important and interesting in their own right but they could also guide future screenings in the rational control of crystal structures and prediction of materials properties.

Here, we focus on nucleation mechanisms during solidification in Ni.
Elemental Ni has been widely used as a model system, for example, for nickel-based superalloys that are used as high temperature materials relevant in the transportation and energy sectors.~\cite{Reed2006}
Consequently, understanding the crystallization processes in these metals is  of great technological importance.  Despite its relevance, even for simple metals like Ni, the microscopic mechanism of crystallization is not completely understood.
Previously we found that structural heterogeneity in the liquid plays an important role during solidification in Ni.~\cite{DLeines2017,DLeines2018} Pre-ordered liquid domains with distinctive bond-orientational order and fcc-hcp like structural features precede the formation of crystallites and act as precursors of the nucleation process, predetermining the polymorph that forms. We have also shown that pre-ordered liquid regions
play a key role in the description of the reaction coordinate and the interfacial free energy~\cite{DLeines2018} and, therefore, are key physical descriptors of the cluster structure and the mechanism. In a first attempt to derive predictive rules for the control of crystallization mechanisms, we have recently demonstrated that the ability of templates to modify the structural features of the liquid, and thus its heterogeneities, is directly linked to the nucleating ability of a template and the promoted polymorphs that emerge.~\cite{DiazLeines2021} Precursor-mediated mechanisms, namely structural heterogeneity, have been shown to play a key role  during the crystallization mechanism of several systems, such as ice, 
hard spheres,~\cite{PhysRevLett.105.025701,PhysRevLett.96.175701} and colloidal models.~\cite{Lechner2011a,Tan2014} 
Recently, a study of ice nucleation has revealed that {\it dynamical} heterogeneity is another key microscopic factor during crystal nucleation, where  extended regions of reduced mobility in the melt precede the emergence of ice crystals, challenging the classical picture of diffusive single particle attachment to a growing crystal cluster.~\cite{Fitzner2019} It was also shown that relatively immobile regions are
composed of structural hallmarks that resemble the final polymorph selected during ice nucleation, showing clear correlation between structural and dynamical precursors during crystallization.
In metallic liquids, it was shown that mobile atoms near the surface of nanowires promote crystallization by allowing a much more effective collective formation of crystal clusters.~\cite{zhang_spatially_2018} Dynamical arrest and heterogeneity have also been found to be the hallmarks of glass transitions~\cite{doi:10.1146/annurev.physchem.51.1.99,Garrahan4701} where spatial correlations in the liquid emerge in terms of motion but not in structure. Yet, it remains to be understood how pre-ordered clusters in the liquid are correlated with regions of different mobility, in connection with the nucleation mechanism, and under which conditions higher or lower mobility yields crystallization or amorphisation.
Given the significant role of structural heterogeneities in the crystallization and glass formation mechanisms in metals, the dynamical  behaviour of the supercooled liquid in nickel alloys 
is bound to be a relevant factor in the nucleation mechanism. How the mobility of the liquid impacts the nucleation mechanism in Ni is, however, unknown. In this work, by analysing a statistical ensemble of crystallization trajectories, obtained from transition path sampling simulations, we show that crystal nucleation in Ni occurs in extended domains of lowest mobility in the supercooled liquid. The regions of low mobility are spatially correlated with the emergence of structural precursors in the liquid that successfully yield critical fluctuations. Furthermore, we illustrate that a collective drop in mobility in the supercooled liquid precedes the formation of pre-ordered regions, indicating that dynamical arrest in the liquid plays a key role in facilitating pre-ordering and subsequent crystallization. Our results demonstrate that dynamical and structural heterogeneity in the liquid are key microscopic factors that promote crystal nucleation events. We expect that future investigations of the decoupling between dynamical arrest and structural pre-ordering in the supercooled liquid will be crucial to understand and predict the nucleating and glass forming ability of other metallic systems. 

\section{Simulation details and calculation of dynamical and structural properties}

\subsection{Computational setup}
All simulations were performed for 3D periodic systems containing $N = 8788$ Ni atoms.  The interactions between Ni atoms were modelled by an embedded atom model (EAM) potential.~\cite{Foiles1986}
The melting temperature of Ni for this potential is $T_m = 1710$~K~\cite{Hoyt2009} and we investigate nucleation at 20~\% undercooling corresponding to $T = 1370$~K.  All molecular dynamics (MD) simulations were performed with the programme package \textsc{lammps}~\cite{Plimpton1995} using a time step of $\Delta t = 2$~fs, and the temperature and pressure were controlled by a Nos{\'e}-Hoover thermostat and barostat.

\subsection{Dynamical heterogeneity}
The relaxation dynamics in the supercooled liquid can be investigated by determining the self-intermediate scattering function~\cite{Berthier2011}
\begin{equation}
\label{eq:scatterf}
F(\mathbf{q},t) = \langle \Phi(\mathbf{q},t) \rangle
\end{equation}
with 
\begin{equation}
\label{eq:phiqt}
\Phi(\mathbf{q},t) = \frac{1}{N} \sum_{j=1}^N \exp\left(i\mathbf{q} \cdot [\mathbf{r}_j(t) - \mathbf{r}_j(0)] \right) \quad ,
\end{equation}
where $N$ is the number of particles, $\mathbf{r}(t)$ is the position of a particle at time $t$, and $\mathbf{q}$ is a reciprocal space vector.
The dynamical heterogeneity (DH) in the supercooled liquid can be further characterised by the dynamical susceptibility $\chi_4(\mathbf{q},t)$~\cite{Donati2002,Toninelli2005}
\begin{equation}
\label{eq:chi4}
\chi_4(\mathbf{q},t) = N \left[ \langle |\Phi(\mathbf{q},t)|^2 \rangle - \langle \Phi(\mathbf{q},t) \rangle^2 \right] \quad .
\end{equation}
In an isotropic system, $F$ and $\chi_4$ can be evaluated for a given value $q_0$ as the average over independent directions, with $F(q_0,t) = \langle F(\mathbf{q},t) \rangle_{||\mathbf{q}|| = q_0}$  and $\chi_4(q_0,t) = \langle \chi_4(\mathbf{q},t) \rangle_{||\mathbf{q}|| = q_0}$, respectively. 
The choice of $q_0$ controls the length scale at which the dynamical processes are observed.  Here, we set $q_0$ to the first peak of the isotropic structure factor
\begin{equation}
\label{eq:strucfac}
S(q) = 1 + \frac{4 \pi \rho}{q} \int_0^{\infty} dr \, r \sin(rq) [g(r)-1] \quad ,
\end{equation}
where $\rho$ is the density of the liquid and $g(r)$ the radial distribution function.
In liquid Ni at 20\% undercooling,  our simulations result in a value of $q_0 = 3.075$, correspondingly.  
We find that the supercooled liquid exhibits only weak DH with an almost exponential relaxation of $F(q_0,t)$.  The time of maximum heterogeneity $t_0$, where the dynamics in the range of nearest neighbours are most heterogeneous, corresponds to the time where $\chi_4(q_0,t)$ has a maximum.  For the investigated system, the corresponding value is $t_0 = 2.64$~ps at the level of supercooling examined. 

Spatially resolved information about the DH is obtained by evaluating the dynamical propensity (DP)~\cite{Sosso2014,Fitzner2019} of each particle
\begin{equation}
\label{eq:dp}
\text{DP}_i (t_0) = \left\langle\frac{|| \mathbf{r}_i(t_0) - \mathbf{r}_i(0) ||^2}{\text{MSD}}\right\rangle_{\text{ISO}} \quad.
\end{equation}
The average is taken over the so-called isoconfigurational ensemble,~\cite{WidmerCooper2007,Colombo2013,Fitzner2019} where a number of MD trajectories is initiated from the same configuration with random velocities drawn from the Maxwell-Boltzmann distribution for a given temperature.  $\text{MSD}$ denotes the mean squared displacement for time $t_0$.  DP values $< 1.0$ indicate particles that are less mobile than the average, and $\text{DP} > 1.0$ for particles that are more mobile, respectively.

\subsection{Structural characterisation}
\label{sec:structure}
To identify solid- and liquid-like particles in the simulations, the criterion introduced by ten Wolde and Frenkel~\cite{Steinhardt1983,Auer2005} was used   A bond between two particles $i$ and $j$ is characterised as solid if $s_{ij} = \sum_{m=-6}^{6} q_{6m}(i) q^{*}_{6m}(j) > 0.5$, where $q_{6m}$ are the complex vectors given by the sum over spherical harmonics with $l=6$.  In addition, the average correlation over neighbouring atoms, $\langle s_i \rangle = 1/N_{\text{nn}} \sum s_{ij}$, was considered to improve the definition of solid particles at the interfaces between solid clusters and the liquid.  A particle $i$ with  at least $7$ solid bonds and $\langle s_i \rangle > 0.6$ is identified as solid. 
The size of the largest cluster composed of solid particles is denoted by $n_s$.

The local structure around each particle was characterised using the averaged local bond-order parameters~\cite{Lechner2008} $\bar{q}_4$ and $\bar{q}_6$.  Details concerning the computation of the corresponding $\bar{q}_4,\bar{q}_6$ reference map for face-centred cubic (fcc), body-centred cubic (bcc), hexagonal closed-packed (hcp), and liquid Ni at 20~\% undercooling can be found in our previous publication.~\cite{DLeines2017}

\section{Results}

\subsection{Supercooled liquid {Ni} exhibits mild dynamical heterogeneity}

\begin{figure} 
\centering
  \includegraphics[width=0.5\textwidth]{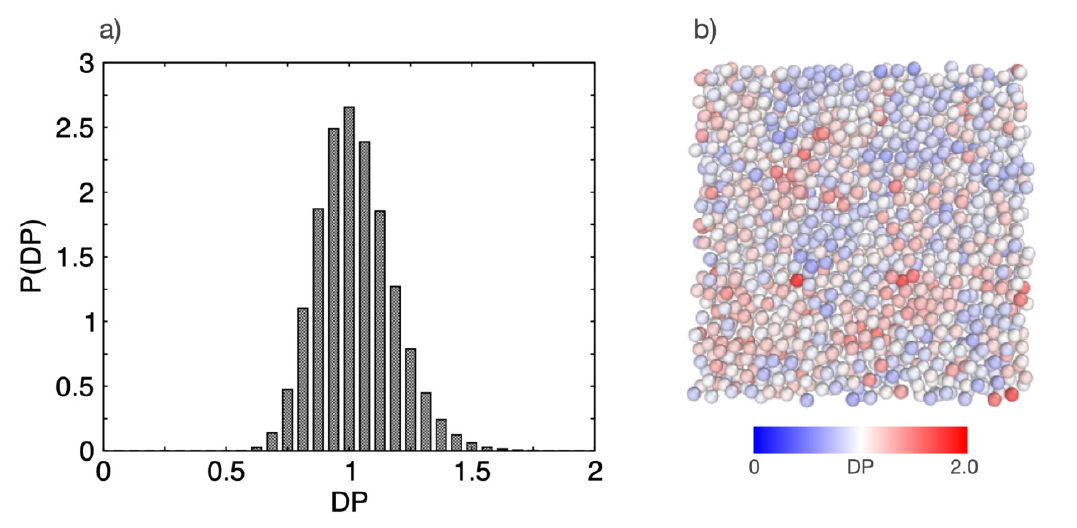}
  \caption{  
  \label{fig:dp_liq}
  {\it Left}: a) Probability distribution of DP values in liquid Ni at 20~\% underccooling; the distribution is centred around $\text{DP} = 1$ and fairly narrow.  
  {\it Right}: b) Snapshot of a representative configuration in the supercooled liquid; atoms are coloured according to their DP value indicating  high (red) and low (blue) mobility.
  }
\end{figure}

\label{sec:liquid}
A long MD trajectory of 10~ns was run in the supercooled liquid at 20~\% undercooling to equilibrate the systems.  From this trajectory, 200 configurations were randomly selected to compute the dynamical propensity of particles in the liquid.  For each configuration, 100 MD simulations with $t_0 = 2.64$~ps were performed to calculate the ensemble average in Eq.~\eqref{eq:dp}.

The probability distribution of DP values in the supercooled liquid is shown in Fig.~\ref{fig:dp_liq}.  The distribution is peaked around $\text{DP} = 1$ and fairly narrow, reflecting the rather small degree of dynamical heterogeneity in the liquid at 20\% undercooling.  Similar to previous studies,~\cite{Sosso2014,Fitzner2019} the top and bottom 5~\% of
the distribution are identified as most mobile (MM, DP > 1.236) and most immobile (MI, DP < 0.728) particles, respectively.  Together with the probability distribution, a representative configuration of the supercooled liquid is shown in Fig.~\ref{fig:dp_liq} where the atoms are coloured according to their DP values.  The spatial distribution of fast (red) and slow (blue) atoms appears mostly random and there is no strong indication of low mobility domains, as expected for a liquid with little dynamical heterogeneity. Our analysis of larger undercoolings, near the nucleation temperature ($\Delta T \sim 30$\%), showed similar behaviour with a slight broadening of the distribution of DP values, indicating that the magnitude of dynamical heterogeneity is not remarkably increased with supercooling. The mild dynamical heterogeneity in supercooled liquid Ni is probably associated with the poor glass forming ability of elemental metals. In contrast, binary alloys and other glass forming materials have been shown to exhibit significant dynamical heterogeneity in the liquid, where MI regions yield dynamical arrest and suppress the emergence of structural precursors of crystallization.~\cite{PhysRevX.8.021040,Sosso2014} Nevertheless, regions of varying mobility (MI  and MM) can be distinguished from random fluctuations in the liquid at 20\% undercooling (Fig.~\ref{fig:dp_liq} b). In the next section we further analyse the correlation of MM and MI regions to structural precursors in the liquid, in connection to the nucleation mechanism during solidification in Ni. 

\subsection{Structural pre-ordering without a decrease in mobility does not promote nucleation}

\begin{figure*} 
\centering
  \includegraphics[width=0.95\textwidth]{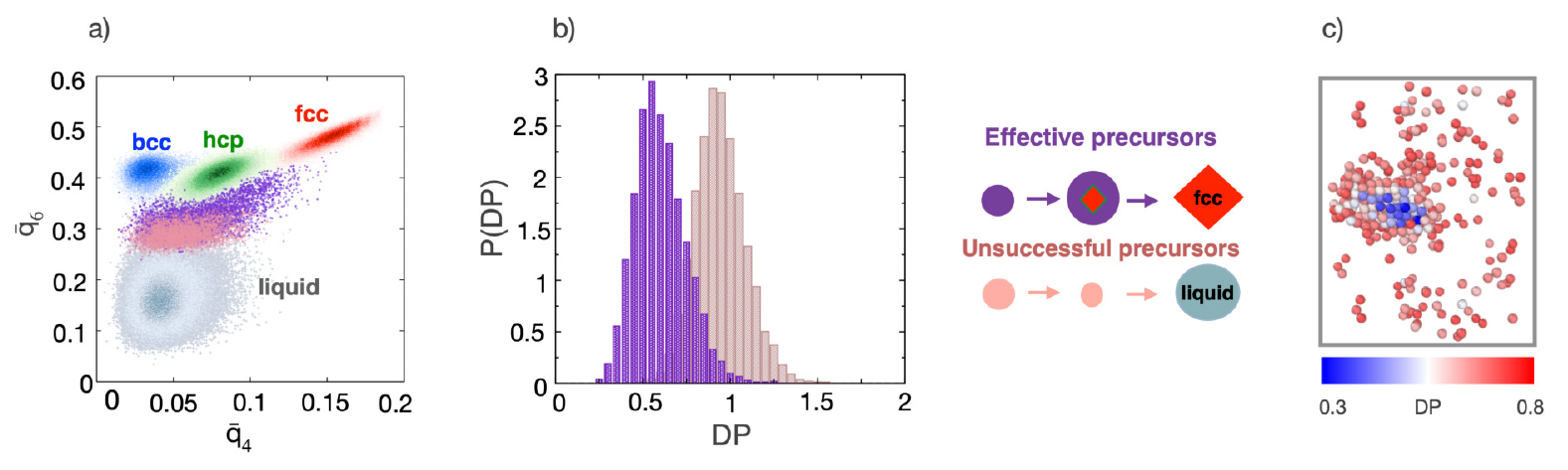}
  \caption{  
  \label{fig:dp_succusucc}
  {\it Left}: a) Map showing the distribution of averaged local bond-order parameters $\bar{q}_4$ and $\bar{q}_6$ for the crystalline bulk phases fcc (red), hcp (green), and bcc (blue), as well as the liquid (grey), together with the values for pre-critical clusters that successfully continue to grow (purple) or dissolve again (pink). Figure adapted from Ref.~\cite{DiazLeines2021}.
  {\it Middle}: b) Probability distribution of DP values for effective (purple) and unsuccessful (pink) precursors; the distribution for unsuccessful precursor is comparable to the one of the liquid shown in Fig.~\ref{fig:dp_liq}a), while the one of effective precursors is clearly shifted to lower DP values.
  {\it Right}: c) Representative configuration of an effective precursor; only atoms with DP < 0.728 (most immobile) are shown; atoms are coloured according to their DP value.
  }
\end{figure*}

Using transition path sampling (TPS)~\cite{Dellago2002,VanErp2005,VanErp2007,Bolhuis2008} simulations,  we established in our previous work~\cite{DLeines2017,DLeines2018} that homogeneous nucleation in Ni proceeds via the initial formation of precursors from which the crystalline phase emerges.  These precursors are composed of solid-like particles that have higher bond-orientational order than the liquid but less than any of the crystalline phases.  This pre-ordering in the liquid is associated with a significant contribution to the nucleation barrier ($\approx 1$~eV at 20~\% undercooling with a barrier of $\Delta G= 4.21$ eV) and plays a crucial role in the reaction coordinate describing the nucleation mechanism.~\cite{DLeines2018}
In addition, we have recently shown that the efficiency of precursors to facilitate the formation of a crystalline bulk phase strongly depends on their structural characteristics.~\cite{DiazLeines2021}  
Here, we focus on pre-critical clusters with $n_s = 50$ that are predominantly ($> 90$~\%) composed of pre-structured liquid particles.  The distribution of $\bar{q}_4,\bar{q}_6$ values of the pre-critical clusters is shown in the left graph of Fig.~\ref{fig:dp_succusucc} (purple and pink) together with the distributions for bulk fcc (red), hcp (green), bcc (blue), and liquid (grey).  The clusters have been extracted from 400 trajectories of the transition path ensemble (TPE) of homogeneous nucleation in Ni (details concerning the TPS simulations are given in Refs.~\cite{DLeines2017,DLeines2018}).
As indicated above, pre-ordered liquid particles exhibit bond-orientational order that differs from both the liquid and the crystalline phases.
The purple distribution represents pre-critical clusters from trajectories that successfully nucleate a crystalline phase and continue to grow, whereas the pink distribution corresponds to clusters that eventually dissolve.  Effective precursors (purple) that successfully initiate the emergence of the crystalline phase and grow beyond the critical nucleus size show a clear increase in bond-orientational order compared to  unsuccessful pre-critical clusters (pink) that shrink and dissolve.

The dynamical propensity of atoms in the pre-critical clusters was obtained 
by selecting 200 configurations with $n_s = 50$ from the TPE of our previous study~\cite{DLeines2017,DLeines2018} and running 100 MD simulations with $t_0 = 2.64$~ps for each configuration to calculate the ensemble average in Eq.~\eqref{eq:dp}. 
The corresponding probability distributions of DP values are shown in the middle graph of Fig.~\ref{fig:dp_succusucc}.  Atoms comprising effective precursors show a strong decrease in DP compared to the liquid, whereas the DP distribution of atoms in unsuccessful precursors is very similar to the one of the liquid shown in Fig.~\ref{fig:dp_liq}a).  
This is rather interesting since these particles are characterised as solid according to the criterion outlined in section~\ref{sec:structure} and, correspondingly, exhibit a higher bond-orientational order than the liquid evidenced by their $\bar{q}_4,\bar{q}_6$ distribution.  Still, their dynamical properties are comparable to those of liquid particles.

Comparing the structural and dynamical properties of effective and unsuccessful precursors, it appears that the structural and dynamical heterogeneity affect the nucleation process differently.  While both effective and unsuccessful precursors show an increase in bond-orientational order compared to the liquid, the DP is reduced only for particles in effective structural precursors that yield critical fluctuations.  Structural pre-ordering in the liquid without a decrease in mobility, therefore, does not promote the nucleation of the crystalline bulk phase.
A snapshot of a configuration with an effective precursor is shown on the right in Fig.~\ref{fig:dp_succusucc}.  The atoms are coloured according to their DP value and only atoms with $\text{DP} < 0.728$ (most immobile) are shown.  These particles clearly form a region of lowest mobility in which the largest pre-structured cluster that succeeds to nucleate is embedded.  

\subsection{Structural pre-ordering emerges in low mobility regions}
\begin{figure*} 
\centering
  \includegraphics[width=0.9\textwidth]{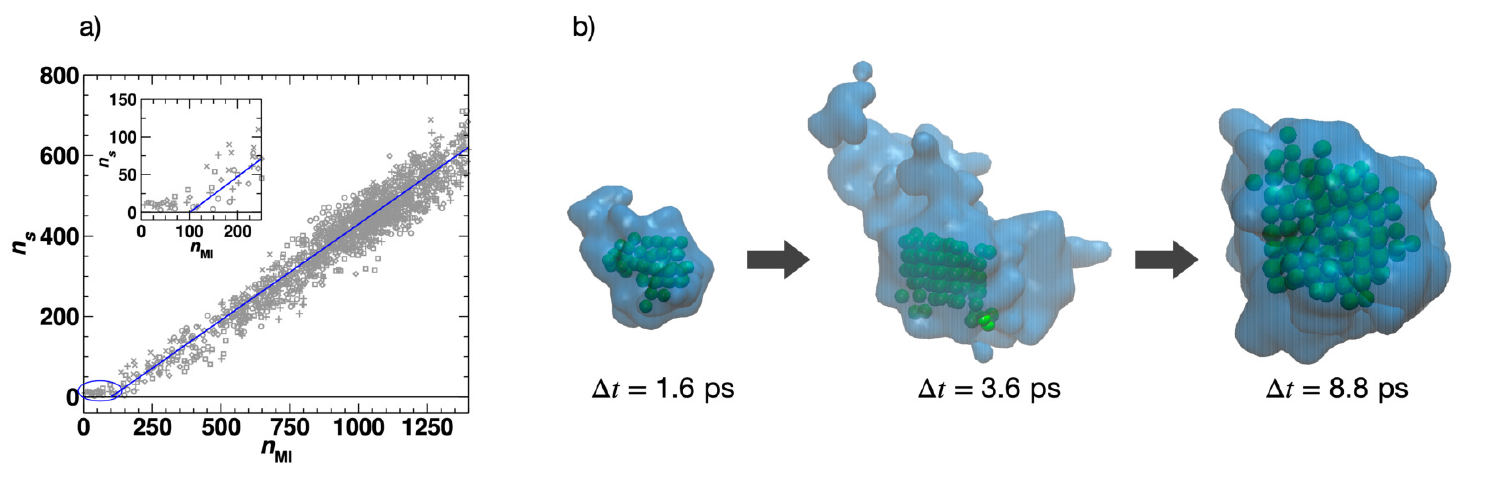}
  \caption{{\it Left}: a) Correlation between the largest solid cluster, $n_s$, and the largest cluster of most immobile particles, $n_{\text{MI}}$.  Up to $n_{\text{MI}} \approx 100$, $n_s$ is approximately constant (marked by the blue ellipse) before it increases linearly with $n_{\text{MI}}$.  The data are extracted from five representative $AB$-trajectories of the TPE (marked by various symbols).  The blue line shows a linear fit to the data. 
  {\it Right}: b) Three configurations along a representative $AB$-trajectory; the blue, transparent region denotes the largest MI cluster, $n_{\text{MI}}$, the green spheres show the largest solid cluster, $n_s$; the time below each configuration corresponds to the transition time, that is the time after leaving the stable liquid state.
  \label{fig:ns_ndp}
  }
\end{figure*}

To follow the evolution of low mobility regions during nucleation, that is regions comprised of the MI particles, we have analysed 100 $AB$-paths (trajectories that successfully crystallise from the liquid state $A$ to the solid state $B$) from the TPE of our previous study.~\cite{DLeines2017,DLeines2018}  In addition to the largest solid cluster $n_s$, the largest cluster composed of MI particles $n_{\text{MI}}$ is determined for each time slice in the $AB$-trajectories.  
In Fig.~\ref{fig:ns_ndp}a), the number of particles in the largest solid cluster is plotted as a function of the number of particles in the largest MI cluster for five representative trajectories.  Up to $n_{\text{MI}} \approx 100$, the size of the $n_s$ is small and remains roughly constant, indicating that the structural pre-ordering in the supercooled liquid is preceded by a distinctive dynamical arrest of spatial regions in the supercooled liquid and a decoupling of the dynamical and structural properties of the melt during induction time.  For larger sizes of $n_{\text{MI}}$, $n_s$ increases linearly with $n_{\text{MI}}$, showing a strong correlation of the MI regions with structural fluctuations that yield crystallization.  The regions of low mobility are always larger than the largest solid clusters, and for stable clusters that continue to grow and crystallize, the overlap between particles that belong to both $n_s$ and $n_{\text{MI}}$ is $> 90$~\%, showing that crystallites nucleate in regions of lowest mobility in the supercooled liquid. Consequently, the low mobility regions encompass the regions of increased structural ordering.  The linear correlation between $n_s$ and $n_{\text{MI}}$ with a slope of $n_{\text{MI}}/n_s \approx 2$ suggests that, within a spherical approximation, the radius of the low mobility region is $\approx 25$~\% larger than the radius of the largest solid cluster.

The largest MI cluster is visualised together with the largest solid cluster for three configurations along a representative $AB$-trajectory in Fig.~\ref{fig:ns_ndp}b).  
The times in Fig.~\ref{fig:ns_ndp}b) correspond to the transition time without the residence time in the stable liquid state, since trajectories in the TPE correspond only to the actual transition itself, while the stable states are sampled separately.
It is clearly visible that the largest solid cluster (green spheres) is located within the regions of low mobility (blue transparent) at different times along the nucleation process.  The behaviour depicted in Fig.~\ref{fig:ns_ndp}b) is typical of all trajectories in the ensemble.  

\section{Conclusions}
We have analysed the structural and dynamical properties of atoms during homogeneous nucleation in nickel.  Our results reveal a clear correlation between the formation of low mobility regions in the liquid and the formation of structural precursors that precede the emergence of the crystalline bulk phase.  Our findings are similar to the results reported for the nucleation of ice,~\cite{Fitzner2019} even though Ni, as an elemental metal, exhibits much less dynamical heterogeneity in the supercooled liquid.  Still, we find that the dynamical propensity of the atoms plays a key role in the initial formation of structurally pre-ordered regions in the liquid.

Most notably, the distribution of DP values for successful precursors that continue to grow and crystallise into the bulk phase is significantly shifted compared to the distribution of DP values for liquid particles.  In contrast, solid clusters that dissolve have DP values similar to the average value of the liquid.
The dynamical properties are thus crucial in the description of the nucleation mechanism.
As the solid clusters continue to grow, they are always embedded in a region of low mobility and atoms in the vicinity of the solid cluster exhibit clearly  decreased DP values.

Understanding the role of dynamical heterogeneity in  nucleation mechanisms has implications that go beyond the description of homogeneous nucleation.  In particular, our findings further support the idea that in heterogeneous nucleation, structural templating is not necessarily the only decisive factor, but the efficiency of nucleating agents might also be determined by their ability to modify the dynamical properties of the liquid.  
Including dynamical heterogeneity in the study of nucleation mechanisms, for both homogeneous and heterogeneous nucleation, provides an additional perspective on the underlying atomistic processes that are key to control central parameters during crystallisation, including nucleation rates and polymorph selectivity.

\section*{Author Contributions}
All authors contributed to the conceptualisation of the study.  GDL and JR  performed the investigation, the formal analysis and visualisation of the research data, and wrote the original draft.  All authors reviewed and edited the final draft.

\section*{Conflicts of interest}
There are no conflicts of interest to declare.

\section*{Acknowledgements}
The authors would like to thank Martin Fitzner for fruitful discussions and support in preparing the simulations for the dynamical heterogeneity of the liquid.
JR acknowledges financial support from the Deutsche Forschungsgemeinschaft (DFG) through the Heisenberg Programme project 428315600. GDL acknowledges support from Conacyt-Mexico through fellowship Ref. No. 220644. GDL and AM acknowledge support from the Isaac Newton Trust Grant Ref No: 20.40(h). The authors acknowledge computing time from UKCP ( EP/P022561/1) and the UK Materials and Molecular Modelling Hub, which is partially funded by EPSRC (EP/P020194/1 and EP/T022213/1)). 


\balance

\providecommand*{\mcitethebibliography}{\thebibliography}
\csname @ifundefined\endcsname{endmcitethebibliography}
{\let\endmcitethebibliography\endthebibliography}{}


\begin{mcitethebibliography}{39}
\providecommand*{\natexlab}[1]{#1}
\providecommand*{\mciteSetBstSublistMode}[1]{}
\providecommand*{\mciteSetBstMaxWidthForm}[2]{}
\providecommand*{\mciteBstWouldAddEndPuncttrue}
  {\def\EndOfBibitem{\unskip.}}
\providecommand*{\mciteBstWouldAddEndPunctfalse}
  {\let\EndOfBibitem\relax}
\providecommand*{\mciteSetBstMidEndSepPunct}[3]{}
\providecommand*{\mciteSetBstSublistLabelBeginEnd}[3]{}
\providecommand*{\EndOfBibitem}{}
\mciteSetBstSublistMode{f}
\mciteSetBstMaxWidthForm{subitem}
{(\emph{\alph{mcitesubitemcount}})}
\mciteSetBstSublistLabelBeginEnd{\mcitemaxwidthsubitemform\space}
{\relax}{\relax}

\bibitem[Sosso \emph{et~al.}(2016)Sosso, Chen, Cox, Fitzner, Pedevilla, Zen,
  and Michaelides]{Sosso2016}
G.~C. Sosso, J.~Chen, S.~J. Cox, M.~Fitzner, P.~Pedevilla, A.~Zen and
  A.~Michaelides, \emph{Chem. Rev.}, 2016, \textbf{116}, 7078--7116\relax
\mciteBstWouldAddEndPuncttrue
\mciteSetBstMidEndSepPunct{\mcitedefaultmidpunct}
{\mcitedefaultendpunct}{\mcitedefaultseppunct}\relax
\EndOfBibitem
\bibitem[Blow \emph{et~al.}(2021)Blow, Quigley, and
  Sosso]{doi:10.1063/5.0055248}
K.~E. Blow, D.~Quigley and G.~C. Sosso, \emph{J. Chem. Phys.}, 2021,
  \textbf{155}, 040901\relax
\mciteBstWouldAddEndPuncttrue
\mciteSetBstMidEndSepPunct{\mcitedefaultmidpunct}
{\mcitedefaultendpunct}{\mcitedefaultseppunct}\relax
\EndOfBibitem
\bibitem[Russo \emph{et~al.}(2018)Russo, Romano, and Tanaka]{PhysRevX.8.021040}
J.~Russo, F.~Romano and H.~Tanaka, \emph{Phys. Rev. X}, 2018, \textbf{8},
  021040\relax
\mciteBstWouldAddEndPuncttrue
\mciteSetBstMidEndSepPunct{\mcitedefaultmidpunct}
{\mcitedefaultendpunct}{\mcitedefaultseppunct}\relax
\EndOfBibitem
\bibitem[Gebauer \emph{et~al.}(2014)Gebauer, Kellermeier, Gale, Bergström, and
  Cölfen]{Gebauer2014}
D.~Gebauer, M.~Kellermeier, J.~D. Gale, L.~Bergström and H.~Cölfen,
  \emph{Chem. Soc. Rev.}, 2014, \textbf{43}, 2348--2371\relax
\mciteBstWouldAddEndPuncttrue
\mciteSetBstMidEndSepPunct{\mcitedefaultmidpunct}
{\mcitedefaultendpunct}{\mcitedefaultseppunct}\relax
\EndOfBibitem
\bibitem[ten Wolde and Frenkel(1997)]{tenWolde1997}
P.~R. ten Wolde and D.~Frenkel, \emph{Science}, 1997, \textbf{277},
  1975--1978\relax
\mciteBstWouldAddEndPuncttrue
\mciteSetBstMidEndSepPunct{\mcitedefaultmidpunct}
{\mcitedefaultendpunct}{\mcitedefaultseppunct}\relax
\EndOfBibitem
\bibitem[ten Wolde and Frenkel(1999)]{tenWolde1999}
P.~R. ten Wolde and D.~Frenkel, \emph{Phys. Chem. Chem. Phys.}, 1999,
  \textbf{1}, 2191--2196\relax
\mciteBstWouldAddEndPuncttrue
\mciteSetBstMidEndSepPunct{\mcitedefaultmidpunct}
{\mcitedefaultendpunct}{\mcitedefaultseppunct}\relax
\EndOfBibitem
\bibitem[Zhang and Liu(2007)]{Zhang2007}
T.~Zhang and X.~Y. Liu, \emph{J. Am. Chem. Soc.}, 2007, \textbf{129},
  13520\relax
\mciteBstWouldAddEndPuncttrue
\mciteSetBstMidEndSepPunct{\mcitedefaultmidpunct}
{\mcitedefaultendpunct}{\mcitedefaultseppunct}\relax
\EndOfBibitem
\bibitem[Prestipino \emph{et~al.}(2012)Prestipino, Laio, and
  Tosatti]{PhysRevLett.108.225701}
S.~Prestipino, A.~Laio and E.~Tosatti, \emph{Phys. Rev. Lett.}, 2012,
  \textbf{108}, 225701\relax
\mciteBstWouldAddEndPuncttrue
\mciteSetBstMidEndSepPunct{\mcitedefaultmidpunct}
{\mcitedefaultendpunct}{\mcitedefaultseppunct}\relax
\EndOfBibitem
\bibitem[Russo and Tanaka(2012)]{Russo2012}
J.~Russo and H.~Tanaka, \emph{Sci. Rep.}, 2012, \textbf{2}, 505--512\relax
\mciteBstWouldAddEndPuncttrue
\mciteSetBstMidEndSepPunct{\mcitedefaultmidpunct}
{\mcitedefaultendpunct}{\mcitedefaultseppunct}\relax
\EndOfBibitem
\bibitem[Fitzner \emph{et~al.}(2019)Fitzner, Sosso, Cox, and
  Michaelides]{Fitzner2019}
M.~Fitzner, G.~C. Sosso, S.~J. Cox and A.~Michaelides, \emph{Proc. Natl. Acad.
  Sci. USA}, 2019, \textbf{116}, 2009--2014\relax
\mciteBstWouldAddEndPuncttrue
\mciteSetBstMidEndSepPunct{\mcitedefaultmidpunct}
{\mcitedefaultendpunct}{\mcitedefaultseppunct}\relax
\EndOfBibitem
\bibitem[D\'{i}az~Leines and Rogal(2018)]{DLeines2018}
G.~D\'{i}az~Leines and J.~Rogal, \emph{J. Phys. Chem. B}, 2018, \textbf{122},
  10934--10942\relax
\mciteBstWouldAddEndPuncttrue
\mciteSetBstMidEndSepPunct{\mcitedefaultmidpunct}
{\mcitedefaultendpunct}{\mcitedefaultseppunct}\relax
\EndOfBibitem
\bibitem[Tanaka(2012)]{Tanaka2012}
H.~Tanaka, \emph{Eur. Phys. J. E}, 2012, \textbf{35}, 113\relax
\mciteBstWouldAddEndPuncttrue
\mciteSetBstMidEndSepPunct{\mcitedefaultmidpunct}
{\mcitedefaultendpunct}{\mcitedefaultseppunct}\relax
\EndOfBibitem
\bibitem[D\'{i}az~Leines \emph{et~al.}(2017)D\'{i}az~Leines, Drautz, and
  Rogal]{DLeines2017}
G.~D\'{i}az~Leines, R.~Drautz and J.~Rogal, \emph{J. Chem. Phys.}, 2017,
  \textbf{146}, 154702\relax
\mciteBstWouldAddEndPuncttrue
\mciteSetBstMidEndSepPunct{\mcitedefaultmidpunct}
{\mcitedefaultendpunct}{\mcitedefaultseppunct}\relax
\EndOfBibitem
\bibitem[Zhang \emph{et~al.}(2019)Zhang, Wang, Tang, Wang, Li, Zhou, and
  Wang]{Zhang2019}
Q.~Zhang, J.~Wang, S.~Tang, Y.~Wang, J.~Li, W.~Zhou and Z.~Wang, \emph{Phys.
  Chem. Chem. Phys.}, 2019, \textbf{21}, 4122--4135\relax
\mciteBstWouldAddEndPuncttrue
\mciteSetBstMidEndSepPunct{\mcitedefaultmidpunct}
{\mcitedefaultendpunct}{\mcitedefaultseppunct}\relax
\EndOfBibitem
\bibitem[Lechner \emph{et~al.}(2011)Lechner, Dellago, and
  Bolhuis]{Lechner2011a}
W.~Lechner, C.~Dellago and P.~G. Bolhuis, \emph{Phys. Rev. Lett.}, 2011,
  \textbf{106}, 085701\relax
\mciteBstWouldAddEndPuncttrue
\mciteSetBstMidEndSepPunct{\mcitedefaultmidpunct}
{\mcitedefaultendpunct}{\mcitedefaultseppunct}\relax
\EndOfBibitem
\bibitem[Tan \emph{et~al.}(2014)Tan, Xu, and Xu]{Tan2014}
P.~Tan, N.~Xu and L.~Xu, \emph{Nat. Phys..}, 2014, \textbf{10}, 73\relax
\mciteBstWouldAddEndPuncttrue
\mciteSetBstMidEndSepPunct{\mcitedefaultmidpunct}
{\mcitedefaultendpunct}{\mcitedefaultseppunct}\relax
\EndOfBibitem
\bibitem[Schilling \emph{et~al.}(2010)Schilling, Sch\"ope, Oettel, Opletal, and
  Snook]{PhysRevLett.105.025701}
T.~Schilling, H.~J. Sch\"ope, M.~Oettel, G.~Opletal and I.~Snook, \emph{Phys.
  Rev. Lett.}, 2010, \textbf{105}, 025701\relax
\mciteBstWouldAddEndPuncttrue
\mciteSetBstMidEndSepPunct{\mcitedefaultmidpunct}
{\mcitedefaultendpunct}{\mcitedefaultseppunct}\relax
\EndOfBibitem
\bibitem[Sch\"ope \emph{et~al.}(2006)Sch\"ope, Bryant, and van
  Megen]{PhysRevLett.96.175701}
H.~J. Sch\"ope, G.~Bryant and W.~van Megen, \emph{Phys. Rev. Lett.}, 2006,
  \textbf{96}, 175701\relax
\mciteBstWouldAddEndPuncttrue
\mciteSetBstMidEndSepPunct{\mcitedefaultmidpunct}
{\mcitedefaultendpunct}{\mcitedefaultseppunct}\relax
\EndOfBibitem
\bibitem[Reed(2006)]{Reed2006}
R.~C. Reed, \emph{The {Superalloys}: {Fundamentals} and {Applications}},
  Cambridge University Press, Cambridge, 2006\relax
\mciteBstWouldAddEndPuncttrue
\mciteSetBstMidEndSepPunct{\mcitedefaultmidpunct}
{\mcitedefaultendpunct}{\mcitedefaultseppunct}\relax
\EndOfBibitem
\bibitem[D\'{i}az~Leines and Rogal(2021)]{DiazLeines2021}
G.~D\'{i}az~Leines and J.~Rogal, \emph{arXiv:2106.00147}, 2021\relax
\mciteBstWouldAddEndPuncttrue
\mciteSetBstMidEndSepPunct{\mcitedefaultmidpunct}
{\mcitedefaultendpunct}{\mcitedefaultseppunct}\relax
\EndOfBibitem
\bibitem[Zhang \emph{et~al.}(2018)Zhang, Maldonis, Liu, Schroers, and
  Voyles]{zhang_spatially_2018}
P.~Zhang, J.~J. Maldonis, Z.~Liu, J.~Schroers and P.~M. Voyles, \emph{Nature
  Communications}, 2018, \textbf{9}, 1129\relax
\mciteBstWouldAddEndPuncttrue
\mciteSetBstMidEndSepPunct{\mcitedefaultmidpunct}
{\mcitedefaultendpunct}{\mcitedefaultseppunct}\relax
\EndOfBibitem
\bibitem[Ediger(2000)]{doi:10.1146/annurev.physchem.51.1.99}
M.~D. Ediger, \emph{Ann. Rev. Phys. Chem.}, 2000, \textbf{51}, 99\relax
\mciteBstWouldAddEndPuncttrue
\mciteSetBstMidEndSepPunct{\mcitedefaultmidpunct}
{\mcitedefaultendpunct}{\mcitedefaultseppunct}\relax
\EndOfBibitem
\bibitem[Garrahan(2011)]{Garrahan4701}
J.~P. Garrahan, \emph{Proc. Natl. Acad. Sci. USA}, 2011, \textbf{108},
  4701--4702\relax
\mciteBstWouldAddEndPuncttrue
\mciteSetBstMidEndSepPunct{\mcitedefaultmidpunct}
{\mcitedefaultendpunct}{\mcitedefaultseppunct}\relax
\EndOfBibitem
\bibitem[Foiles \emph{et~al.}(1986)Foiles, Baskes, and Daw]{Foiles1986}
S.~M. Foiles, M.~I. Baskes and M.~S. Daw, \emph{Phys. Rev. B}, 1986,
  \textbf{33}, 7983--7991\relax
\mciteBstWouldAddEndPuncttrue
\mciteSetBstMidEndSepPunct{\mcitedefaultmidpunct}
{\mcitedefaultendpunct}{\mcitedefaultseppunct}\relax
\EndOfBibitem
\bibitem[Hoyt \emph{et~al.}(2009)Hoyt, Olmsted, Jindal, Asta, and
  Karma]{Hoyt2009}
J.~J. Hoyt, D.~Olmsted, S.~Jindal, M.~Asta and A.~Karma, \emph{Phys. Rev. E},
  2009, \textbf{79}, 020601\relax
\mciteBstWouldAddEndPuncttrue
\mciteSetBstMidEndSepPunct{\mcitedefaultmidpunct}
{\mcitedefaultendpunct}{\mcitedefaultseppunct}\relax
\EndOfBibitem
\bibitem[Plimpton(1995)]{Plimpton1995}
S.~Plimpton, \emph{J. Comp. Phys.}, 1995, \textbf{117}, 1--19\relax
\mciteBstWouldAddEndPuncttrue
\mciteSetBstMidEndSepPunct{\mcitedefaultmidpunct}
{\mcitedefaultendpunct}{\mcitedefaultseppunct}\relax
\EndOfBibitem
\bibitem[Berthier and Biroli(2011)]{Berthier2011}
L.~Berthier and G.~Biroli, \emph{Rev. Mod. Phys.}, 2011, \textbf{83},
  587--645\relax
\mciteBstWouldAddEndPuncttrue
\mciteSetBstMidEndSepPunct{\mcitedefaultmidpunct}
{\mcitedefaultendpunct}{\mcitedefaultseppunct}\relax
\EndOfBibitem
\bibitem[Donati \emph{et~al.}(2002)Donati, Franz, Glotzer, and
  Parisi]{Donati2002}
C.~Donati, S.~Franz, S.~C. Glotzer and G.~Parisi, \emph{J. Non-Cryst. Solids},
  2002, \textbf{307-310}, 215--224\relax
\mciteBstWouldAddEndPuncttrue
\mciteSetBstMidEndSepPunct{\mcitedefaultmidpunct}
{\mcitedefaultendpunct}{\mcitedefaultseppunct}\relax
\EndOfBibitem
\bibitem[Toninelli \emph{et~al.}(2005)Toninelli, Wyart, Berthier, Biroli, and
  Bouchaud]{Toninelli2005}
C.~Toninelli, M.~Wyart, L.~Berthier, G.~Biroli and J.-P. Bouchaud, \emph{Phys.
  Rev. E}, 2005, \textbf{71}, 041505\relax
\mciteBstWouldAddEndPuncttrue
\mciteSetBstMidEndSepPunct{\mcitedefaultmidpunct}
{\mcitedefaultendpunct}{\mcitedefaultseppunct}\relax
\EndOfBibitem
\bibitem[Sosso \emph{et~al.}(2014)Sosso, Colombo, Behler, Del~Gado, and
  Bernasconi]{Sosso2014}
G.~C. Sosso, J.~Colombo, J.~Behler, E.~Del~Gado and M.~Bernasconi, \emph{J.
  Phys. Chem. B}, 2014, \textbf{118}, 13621--13628\relax
\mciteBstWouldAddEndPuncttrue
\mciteSetBstMidEndSepPunct{\mcitedefaultmidpunct}
{\mcitedefaultendpunct}{\mcitedefaultseppunct}\relax
\EndOfBibitem
\bibitem[Widmer-Cooper and Harrowell(2007)]{WidmerCooper2007}
A.~Widmer-Cooper and P.~Harrowell, \emph{J. Chem. Phys.}, 2007, \textbf{126},
  154503\relax
\mciteBstWouldAddEndPuncttrue
\mciteSetBstMidEndSepPunct{\mcitedefaultmidpunct}
{\mcitedefaultendpunct}{\mcitedefaultseppunct}\relax
\EndOfBibitem
\bibitem[Colombo \emph{et~al.}(2013)Colombo, Widmer-Cooper, and
  Del~Gado]{Colombo2013}
J.~Colombo, A.~Widmer-Cooper and E.~Del~Gado, \emph{Phys. Rev. Lett.}, 2013,
  \textbf{110}, 198301\relax
\mciteBstWouldAddEndPuncttrue
\mciteSetBstMidEndSepPunct{\mcitedefaultmidpunct}
{\mcitedefaultendpunct}{\mcitedefaultseppunct}\relax
\EndOfBibitem
\bibitem[Steinhardt \emph{et~al.}(1983)Steinhardt, Nelson, and
  Ronchetti]{Steinhardt1983}
P.~J. Steinhardt, D.~R. Nelson and M.~Ronchetti, \emph{Phys. Rev. B}, 1983,
  \textbf{28}, 784--805\relax
\mciteBstWouldAddEndPuncttrue
\mciteSetBstMidEndSepPunct{\mcitedefaultmidpunct}
{\mcitedefaultendpunct}{\mcitedefaultseppunct}\relax
\EndOfBibitem
\bibitem[Auer and Frenkel(2005)]{Auer2005}
S.~Auer and D.~Frenkel, \emph{Adv. Polym. Sci.}, 2005, \textbf{173},
  149--207\relax
\mciteBstWouldAddEndPuncttrue
\mciteSetBstMidEndSepPunct{\mcitedefaultmidpunct}
{\mcitedefaultendpunct}{\mcitedefaultseppunct}\relax
\EndOfBibitem
\bibitem[Lechner and Dellago(2008)]{Lechner2008}
W.~Lechner and C.~Dellago, \emph{J. Chem. Phys.}, 2008, \textbf{129},
  114707\relax
\mciteBstWouldAddEndPuncttrue
\mciteSetBstMidEndSepPunct{\mcitedefaultmidpunct}
{\mcitedefaultendpunct}{\mcitedefaultseppunct}\relax
\EndOfBibitem
\bibitem[Dellago \emph{et~al.}(2002)Dellago, Bolhuis, and
  Geissler]{Dellago2002}
C.~Dellago, P.~Bolhuis and P.~L. Geissler, \emph{Adv. Chem. Phys.}, 2002,
  \textbf{123}, 1--78\relax
\mciteBstWouldAddEndPuncttrue
\mciteSetBstMidEndSepPunct{\mcitedefaultmidpunct}
{\mcitedefaultendpunct}{\mcitedefaultseppunct}\relax
\EndOfBibitem
\bibitem[van Erp and Bolhuis(2005)]{VanErp2005}
T.~S. van Erp and P.~G. Bolhuis, \emph{J. Comp. Phys.}, 2005, \textbf{205},
  157--181\relax
\mciteBstWouldAddEndPuncttrue
\mciteSetBstMidEndSepPunct{\mcitedefaultmidpunct}
{\mcitedefaultendpunct}{\mcitedefaultseppunct}\relax
\EndOfBibitem
\bibitem[van Erp(2007)]{VanErp2007}
T.~S. van Erp, \emph{Phys. Rev. Lett.}, 2007, \textbf{98}, 268301\relax
\mciteBstWouldAddEndPuncttrue
\mciteSetBstMidEndSepPunct{\mcitedefaultmidpunct}
{\mcitedefaultendpunct}{\mcitedefaultseppunct}\relax
\EndOfBibitem
\bibitem[Bolhuis(2008)]{Bolhuis2008}
P.~G. Bolhuis, \emph{J. Chem. Phys.}, 2008, \textbf{129}, 114108\relax
\mciteBstWouldAddEndPuncttrue
\mciteSetBstMidEndSepPunct{\mcitedefaultmidpunct}
{\mcitedefaultendpunct}{\mcitedefaultseppunct}\relax
\EndOfBibitem
\end{mcitethebibliography}
\end{document}